\documentclass[11pt,epsf]{article}

\usepackage{epsfig}
\usepackage{amssymb}
\usepackage{graphicx}
\usepackage{color}
\usepackage{subfigure}

\begin{document}

\baselineskip 6mm
\renewcommand{\thefootnote}{\fnsymbol{footnote}}

%------------ Chanyong Park's macro's, etc  -----------

\newcommand{\nc}{\newcommand}
\newcommand{\rnc}{\renewcommand}

%%%%%%%%%%%%%%%%%%%%%% Equation Numbering %%%%%%%%%%%%%%%%%%%%%%%
%\makeatletter \rnc{\theequation}{\thesection.\arabic{equation}}
%\@addtoreset{equation}{section} \makeatother

%%%%%%%%%%%%%%%%%%%%%%%%%%%%%%%%%%%%%%%%%%%%%%%%%%%%%%%%%%%%%%%%%
%                                                               %
%                NEW COMMANDS AND MACROS                        %
%                                                               %
%%%%%%%%%%%%%%%%%%%%%%%%%%%%%%%%%%%%%%%%%%%%%%%%%%%%%%%%%%%%%%%%%

\newcommand{\tcb}{\textcolor{blue}}
\newcommand{\tcr}{\textcolor{red}}
\newcommand{\tcg}{\textcolor{green}}

%%%%% Simplify some frequently used LaTeX commands %%%%%

\def\be{\begin{equation}}
\def\ee{\end{equation}}
\def\ba{\begin{array}}
\def\ea{\end{array}}
\def\bea{\begin{eqnarray}}
\def\eea{\end{eqnarray}}
\def\nn{\nonumber\\}

%%%%%  Temporary notation %%%%

\def\ct{\cite}
\def\la{\label}
\def\eq#1{(\ref{#1})}

%%% Greek letters %%%

\def\a{\alpha}
\def\b{\beta}
\def\g{\gamma}
\def\G{\Gamma}
\def\d{\delta}
\def\D{\Delta}
\def\ep{\epsilon}
\def\e{\eta}
\def\ph{\phi}
\def\Ph{\Phi}
\def\ps{\psi}
\def\Ps{\Psi}
\def\k{\kappa}
\def\l{\lambda}
\def\L{\Lambda}
\def\m{\mu}
\def\n{\nu}
\def\th{\theta}
\def\Th{\Theta}
\def\r{\rho}
\def\s{\sigma}
\def\S{\Sigma}
\def\ta{\tau}
\def\o{\omega}
\def\O{\Omega}
\def\pr{\prime}

%%%%% Mathematical Symbols

\def\half{\frac{1}{2}}

\def\goto{\rightarrow}

\def\na{\nabla}
\def\grad{\nabla}
\def\curl{\nabla\times}
\def\div{\nabla\cdot}
\def\pa{\partial}

\def\bra{\left\langle}
\def\ket{\right\rangle}
\def\lb{\left[}
\def\lc{\left\{}
\def\ls{\left(}
\def\lp{\left.}
\def\rp{\right.}
\def\rb{\right]}
\def\rc{\right\}}
\def\rs{\right)}

\def\vac#1{\mid #1 \rangle}

%%%%  Special symbol

\def\td#1{\tilde{#1}}
\def\check{ \maltese {\bf Check!}}

%%%%% Roman pont in math

\def\Tr{{\rm Tr}\,}
\def\det{{\rm det}}

%%%%% Special format

\def\bc#1{\nnindent {\bf $\bullet$ #1} \\ }
\def\ch {$<Check!>$ }
\def\ss {\vspace{1.5cm}}

\begin{titlepage}

%---------------- preprint number ---------------
\hfill\parbox{5cm} { }

\vspace{25mm}

\begin{center}
%------------------------ title ------------------------
{\Large \bf  Finite size effect on the magnon's correlation functions}

%---------------- authors and addresses ----------------
\vskip 1. cm
  { Bum-Hoon Lee$^{ab}$\footnote{e-mail : bhl@sogang.ac.kr}
  and Chanyong Park$^a$\footnote{e-mail : cyong21@sogang.ac.kr}
  }

\vskip 0.5cm

{\it $^a\,$Center for Quantum Spacetime (CQUeST), Sogang University, Seoul 121-742, Korea}\\
{ \it $^b\,$ Department of Physics,, Sogang University, Seoul 121-742, Korea }\\

\end{center}

\thispagestyle{empty}

\vskip2cm

%----------------------- abstract ----------------------

\centerline{\bf ABSTRACT} \vskip 4mm

\vspace{1cm}
We calculate the finite size correction on the three-point correlation function between two giant magnons and one marginal operator. We also check that the structure constant in the string set-up is exactly the same as one of the RG analysis in the gauge theory.

\vspace{2cm}
%PACS numbers:

%\today

\end{titlepage}

\renewcommand{\thefootnote}{\arabic{footnote}}
\setcounter{footnote}{0}

\tableofcontents
%%%%%%%%%%%%%%%%%%%%%%%%%%%%%%
%                                                                            %
%   Sec.  Introduction                                                       %
%                                                                            %
%%%%%%%%%%%%%%%%%%%%%%%%%%%%%%

\section{Introduction}
The conformal field theory (CFT) is characterized by the conformal dimension
of all primary operators and the structure constant included in the three-point correlation functions,
because higher point functions may be determined by using the operator product expansion (OPE).
The ${\cal N}=4$ super Yang-Mills (SYM) theory in four-dimensional space is an important example to
investigate the interacting CFT \ct{mal1}.  After it was shown that there exists an integrable structure
in ${\cal N}=4$ SYM theory \cite{Beisert:2005tm,Minahan:2002ve,Beisert:2003xu,Beisert:2003yb}, there was a 
great progress in finding the spectrum of this theory \cite{Ishizeki:2007we}-\cite{Cavaglia:2011kd}.
These studies were extended to the
ABJM model corresponding to  the low energy theory of M-theory or IIA-string theory
\cite{Aharony:2008ug}-\cite{Gromov:2009at}.
On the contrary, although the structure constant can be evaluated in the weak coupling limit of
SYM by computing the Feynman diagrams,
at the strong coupling there still remain many things to be done.

After the proposition of the method calculating the three-point correlation function between two heavy operators and one marginal operator \cite{Janik:2010gc}, there were many interesting works calculating two- and three-point correlation function semi-calssically by using the known explicit solutions  \cite{Buchbinder:2010vw}-\cite{Hernandez:2011up}.
From these works, it was checked that two- and three-point correlation function in the string
theory are exactly consistent with the RG calculation in the dual SYM.
Another important property of this dual gauge theory is the wrapping effect \cite{Janik:2006dc,Arutyunov:2006gs}, which provided a clue for the all-loop Bethe ansatz.
The wrapping effect of the spin chain model was studied by investigating the finite
size effect in the dispersion relation of the giant magnon in the string theory.
In this paper, as the generalization of Ref. \cite{Park:2010vs,Bai:2011su},  we are going to  investigate the finite size effect of the three-point correlation function
between two magnons and one marginal operator and then compare this result with the structure constant
of the dual gauge theory obtained by the RG analysis.
Finally, we will show that the finite size effect on the three-point correlation function calculated in the string
theory are exactly matched with ones obtained in the gauge theory, which may provide another
evident example for the AdS/CFT correspondence. 

The rest part is organized as follows. In Sec. 2, we will investigate the finite size correction
on the two-point correlation function of the giant magnon by using the saddle point approximation.
In Sec. 3, we will calculate the finite size effect of the three-point correlation function between two magnons and one marginal operator. Finally, we will finish our work with a brief discussion
in Sec.4.

{\bf Note added} At the final stage of this work, we noticed that there were overlaps in Ref. \cite{Ahn:2011zg},
in which the different method, so called the Neumann-Rosochatius reduction, with the
known solution was used.

\section{Finite size effect on the dispersion relation of the giant magnon}

Consider a solitonic string moving in the $AdS_5 \times S^2$, which is a
subspace of $AdS_5 \times S^5$. In the Euclidean Poincare patch
\be
ds_{AdS}^2 = \frac{1}{z^2}  \ls  dz^2 + d \vec{x}^2 \rs ,
\ee
the string solution can be described by a point-particle moving in $AdS$.
Especially, in the conformal gauge the integration over the string worldsheet is reduced to
the integration over the modular parameter $s$ of the cylinder
\be
\int d^2 \s \to \int_{-s/2}^{s/2}  d \ta \int_{-L}^{L} d\s ,
\ee
where we concentrate on the magnon solution and $\pm L$ imply two ends
of the string worldsheet.  Notice that the giant magnon, which is dual of the magnon on the open
spin chain in the dual gauge theory, is described by the worldsheet solitonic 
solution on the open string in which one should give up the level matching 
condition \cite{Hofman:2006xt,Arutyunov:2006gs}.

The string action on $AdS_5 \times S^2$ is given by \cite{Lee:2008ui,Lee:2008yq}
\bea	\la{act:ph0}
S_{st} &=& \int d^2 \s \ {\cal L}_{st} \nn
&=& - \frac{T}{2} \int d^2 \s \lb - \frac{(\pa_{\ta} x )^2 + (\pa_{\ta} z )^2 }{z^2}- \ls
\pa_{\ta} \th \rs^2 + \ls  \pa_{\s} \th \rs^2
- \sin^2 \th \lc  \ls  \pa_{\ta} \ph \rs^2 -\ls  \pa_{\s} \ph \rs^2 \rc  \rb ,
\eea
where $\frac{T}{2}$ is a string tension,  $T=\frac{\sqrt{\l}}{2 \pi}$ for $AdS_5 \times S^5$.
The solutions of the equations of motion for $AdS$ coordinates, $z(\ta)$ and $x(\ta)$ are given by
\bea	\la{sol:ads}
z(\ta) &=& \frac{R}{\cosh \k \ta} , \nn
x(\ta) &=& R \tanh \k  \ta + x_0 ,
\eea
which is the specific parameterization of a geodesic in $AdS$, $(x(\ta) - x_0)^2 +z(\ta)^2 = R^2$.
From these solutions, the action of the $AdS$ part simplifies to
\be		\la{res:adsac}
 \frac{T}{2} \int_{-s/2}^{s/2} d \ta \int_{-L}^{L} d \s \frac{(\pa_{\ta} x )^2 + (\pa_{\ta} z )^2 }{z^2} =
  \frac{T}{2} \int_{-s/2}^{s/2} d \ta \int_{-L}^{L} d \s \k^2 .
\ee
Imposing the boundary conditions
\be
\ls x(-s/2) , z(-s/2) \rs = (0,\ep) \ \  {\rm and } \ \ \ls x(s/2) , z(s/2) \rs = (x_f,\ep) ,
\ee
in which $\ep$ is very small and corresponds to an appropriate UV cut-off,
we can find a relation between $\k$ and $x_f$
\be	\la{rel:kappa}
\k \approx \frac{2}{s} \log \frac{x_f}{\ep} ,
\ee
with $x_f \approx 2 R \approx 2 x_0$.

Now, consider the equations of motion for $S^2$ coordinates. Under the following parameterization
\be
\th = \th (y) \ , \ \  \ph = \n \ta + g (y) \ , \ \ {\rm and}  \  y = a \ta + b \s ,
\ee
the equations of motion for $\ph$ reads off
\be
0 = \pa_y \lc \sin^2 \th \ls a \n + (a^2 - b^2) g' \rs \rc ,
\ee
where the prime means the derivative with respect to $y$.
So $g'$ can be rewritten
in terms of $\th$ as
\be	\la{eq:ph}
g' = \frac{1}{b^2 - a^2} \ls a \n - \frac{c}{\sin^2 \th }\rs ,
\ee
where $c$ is an integration constant. The equation of motion for $\th$ after multiplying $2 \th'$
can be rewritten as the following form
\be
0 = \pa_y \ls \th'^2 + \frac{b^2 \n^2}{(b^2 - a^2)^2} \sin^2 \th + \frac{c^2}{(b^2-a^2)^2 \sin^2 \th} \rs .
\ee
From the above, we can also rewrite $\th'$ in terms of $\th$
\be 	\la{eq:th}
\th'^2 = \frac{b^2 \n^2}{(b^2 - a^2)^2 \sin^2 \th} \lb - \sin^4 \th +\frac{w^2}{b^2 \n^2} \sin^2 \th
- \frac{c^2}{b^2 \n^2} \rb .
\ee
where $\frac{w^2}{(b^2 - a^2)^2}$ is introduced as another integration constant.
If we solve \eq{eq:ph} and \eq{eq:th}, we should introduce two additional constants, which fix
the position of the giant magnon on $S^2$. The integration constants, $c$ and $w$, in the above
determine the velocity of the giant magnon in $\th$- and $\ph$-directions.
As a result, totally four integration constants appear in the exact solution.  
However, since we are interested in the magnon's
dispersion relation, which is described by the conserved charges including one derivative, the 
additional two integration constants are irrelevant. 

Now, we determine two integrations constants, $c$ and $w$, by imposing the appropriate boundary conditions. First, we impose that $\th$ has a maximum value $\th_{max}$ at which $\th'$ is zero.
In the infinite size limit $\th_{\max} = \pi/2$, this boundary condition makes the giant magnon
have the infinite energy and angular momentum, which is the typical structure of the magnon's
dispersion relation. By imposing this boundary condition, \eq{eq:th} can be rewritten as
\be 
\th'^2 = \frac{b^2 \n^2}{(b^2 - a^2)^2 \sin^2 \th}  \lb \ls \sin^2 \th_{max} - \sin^2 \th \rs
\ls \sin^2 \th - \sin^2 \th_{min} \rs \rb ,
\ee
with
\bea
\sin^2 \th_{max} + \sin^2 \th_{min} &=& \frac{w^2}{b^2 \n^2}  , \nn
\sin^2 \th_{max}  \ \sin^2 \th_{min} &=& \frac{c^2}{b^2 \n^2} .
\eea
Another important structure of the magnon's dispersion relation in the infinite size limit is that the difference between the energy and angular momentum of magnon is finite, which can be achieved by
imposing the second boundary condition, $\pa_{\s} \ph = 0$ at $\th=\th_{max}$.
For the finite size case, we can also apply these two boundary conditions to determine the
magnon's dispersion relation. These two boundary conditions fix two integration constants, $c$ and $w$, as
\be
\sin^2 \th_{max} = \frac{c}{a \n} \quad {\rm and} \quad \sin^2 \th_{min} =  \frac{ac}{b^2 \n} ,
\ee
with
\be
w^2 = \frac{c \n (a^2 +b^2)}{a} .
\ee

Following the prescription  of Ref. \cite{Janik:2010gc}, we should consider the evolution of the wave function. Then, the new action $\bar{S}$ including the convolution with the wave function
is given by
\be
\bar{S} \equiv S - \int d^2 \s \ \Pi_{\th} \dot{\th} - \int d^2 \s \ \Pi_{\ph} \dot{\ph}  \nn
=  - \frac{T}{2}
\int_{-s/2}^{s/2} d \ta \int_{-L}^{L} d \s \ \r^2,
\ee
with
\be
\r \equiv \sqrt{\frac{\n c}{a}} ,
\ee
where $2L$ is the length of the worldsheet string. 
Using this action on $S^2$ together with \eq{res:adsac} on $AdS_5$,
the total action for the magnon
is given by
\be
i S_{tot} \equiv i \ls S_{AdS} + \bar{S} \rs = i \ls \frac{4}{s^2} \log^2 \frac{x_f}{\ep}  - \r^2 \rs s L T .
\ee
From this total action, the saddle point of the modular parameter $s$ reads
\be	\la{res:spmag}
\bar{s} = - i \frac{2}{\r} \log \frac{x_f}{\ep} ,
\ee
which corresponds to the Virasoro constraint for the einbein.
At this saddle point, $\k$ and $\r$ are related by $\k = i \r$ and the semi-classical partition function
of the giant magnon
becomes
\be	\la{res:scpf}
e^{i S_{tot}}  = \ls  \frac{\ep}{x_f} \rs^{2 E} ,
\ee
where $E$ corresponds to the magnon's energy. Notice that the form of the semi-classical partition function for the finite size giant magnon is the same as the result of the infinite one. So, the finite
size effect comes from the definition of the conserved charges. The corresponding conserved charges are given by 
\bea   
E &=& 2T \ \frac{z_{max}^2 - z_{min}^2}{z_{max} \sqrt{1-z_{min}^2}} \ K(x), \la{form:energy} \\
J &=& 2 T z_{max} \  \lb  K(x) - E(x) \rb   , \la{form:angular}\\
\frac{\D \ph}{2} &=& \frac{p}{2} = \frac{\sqrt{1-z_{min}^2}}{z_{max} \sqrt{1-z_{max}^2}} \Pi 
\ls \frac{z_{max}^2 - z_{min}^2}{\sqrt{z_{max}^2-1}};x \rs - \frac{\sqrt{1-z_{max}^2}}{z_{max} \sqrt{1-z_{min}^2}}  K(x) , \la{form:momentum}
\eea
with the elliptic integrals of the first, second and third kinds
 \bea 
K(x) &=&  \int_{z_{min}}^{z_{max}} dz 
     \frac{z_{max}}{\sqrt{(z_{max}^2 - z^2)(z^2 - z_{min}^2)}}  , \nn
E(x) &=&  \int_{z_{min}}^{z_{max}} dz \frac{z^2}{z_{max} 
     \sqrt{(z_{max}^2 - z^2)(z^2 - z_{min}^2)}} , \nn 
\Pi \ls \frac{z_{max}^2 - z_{min}^2}{\sqrt{z_{max}^2-1}};x \rs &=&
     \int_{z_{min}}^{z_{max}} dz \frac{ z_{max} (1-z_{max}^2)}{(1-z^2) 
     \sqrt{(z_{max}^2 - z^2)(z^2- z_{min}^2) }} , 
\eea
where $z=\cos \th$, $z_{max}^2 \equiv \cos^2 \th_{min}$, $z_{min}^2 \equiv \cos^2 \th_{max}$ and $x = \sqrt{1-\frac{z_{min}^2}{z_{max}^2}}$. In the case of the large but finite angular momentum, $J/T \gg 1$, since $z_{min}$ is very small, we can expand the elliptic integrals around the limit $z_{min} \to 0$ 
\bea
K(x) &=& \log \ls 4 \frac{z_{max}}{z_{min}} \rs  + \frac{1}{4} \frac{z_{min}^2}{z_{max}^2} \ls \log \ls  4 \frac{z_{max}}{z_{min}} \rs - 1 \rs + \cdots , \nn
E(x) &=& 1 + \frac{1}{4} \frac{z_{min}^2}{z_{max}^2} \ls 2 \log \ls  4 \frac{z_{max}}{z_{min}} \rs - 1 \rs + \cdots , \nn
\Pi \ls \frac{z_{max}^2 - z_{min}^2}{\sqrt{z_{max}^2-1}};x \rs &=& (1-z_{max}^2) 
\lb \log \ls  4 \frac{z_{max}}{z_{min}} \rs  \rp \nn
&& \lp + \frac{1}{4} \frac{z_{min}^2}{z_{max}^2} \lc 
(2 z_{max}^2 +1) \log \ls  4 \frac{z_{max}}{z_{min}} \rs  - (z_{max}^2 + 1 ) \rc  \rb \nn
&&    +
\ls 1 + \half z_{min}^2 \rs z_{max} \sqrt{1 - z_{max}^2} \arcsin z_{max} + \cdots .
\eea
From the above, $z_{max}$ and $\log \ls  4 \frac{z_{max}}{z_{min}} \rs$ can be rewritten in terms of $J$ and $p$ up to $z_{min}^2$ order as
\bea
z_{max} &=& \sin \frac{p}{2} + \frac{1}{4} \frac{(1-\sin^2 \frac{p}{2})}{\sin \frac{p}{2}} \ls \frac{J}{T \sin \frac{p}{2}} + 3 \rs z_{min}^2 \\
\log \ls  4 \frac{z_{max}}{z_{min}} \rs &=& \frac{J}{2 T \sin \frac{p}{2}} + 1 
+ \lb \frac{1}{4  \sin^2 \frac{p}{2}}  
-  \frac{J }{ 8 T \sin^3 \frac{p}{2}} \ls 2 + \frac{J}{T \sin \frac{p}{2}} - 3 \sin^2 \frac{p}{2} 
- \frac{J  \sin \frac{p}{2}}{T}  \rs \rb z_{min}^2 \la{res:expan} . \nn
\eea
Furthermore, since $z_{min}$ is already the small value, from \eq{res:expan}  it  is given at the leading order by
\be \la{rel:zmin}
z_{min} = 4 \sin \frac{p}{2}  \ e^{- \frac{J}{2 T \sin \frac{p}{2}} } ,
\ee
where the small correction proportional to $T/J$ in the exponent is ignored because we
consider the large angular momentum limit, $J \gg T$.
Using these results, we can write $E-J$ in terms of $J$ and $p$ up to $z_{min}^2$ order
\bea	\la{rel:disp}
E - J &=& 2 T \lb \frac{z_{max}^2 - z_{min}^2}{z_{max} \sqrt{1- z_{min}^2}}  K(x) - z_{max}
\lc K(x) - E(x) \rc \rb \nn
&=& 2 T \sin \frac{p}{2} - 8 T \sin^3 \frac{p}{2} \ e^{- \frac{J}{T \sin \frac{p}{2}}} ,
\eea
which is the same as the result in Ref. \cite{Arutyunov:2006gs}. If we take the limit $J \to \infty$, we can ignore the second term, so the above is reduced to the dispersion relation of the giant magnon in the infinite size limit.

\section{Finite size effect on the three-point correlation function}

In this section,  we will investigate the finite size effect on the three-point correlation function between two magnons and one marginal
scalar operator. To calculate this correlation function in the string set-up, we introduce a massless scalar field which is dual to the marginal scalar operator.  
The bulk-to-boundary propagator of a massless scalar field $\chi$ in $AdS$ is given by
\cite{Freedman:1998tz}
\be
K_{\chi} (x^{\m},z;y^{\n}) = \frac{6}{\pi^2 } \ls \frac{z}{z^2 + (x-y)^2 } \rs^4 .
\ee
Then, the three-point function  between two magnon operators denoted by  ${\cal  O}_m$ and marginal
scalar operator ${\cal D}_{\chi}$ is given by \cite{Costa:2010rz}
\be	\la{for:three}
\bra {\cal O}_m (0) {\cal O}_m (x_f)  {\cal D}_{\chi} (y) \ket \approx
\frac{I_{\chi}[\bar{X},\bar{s};y]}{|x_f|^{2 E}}  ,
\ee
with
\be
I_{\chi}[X,s;y] = i \int_{-s/2}^{s/2} d\ta \int_{-L}^{L} d \s \lp \frac{\d S_p[X,s,\chi]}{\d \chi}
\right|_{\chi=0} \ K_{\chi} \ls X(\ta,\s);y \rs ,
\ee
where $\chi$ corresponds to the massless dilaton fluctuation and $S_p[X,s,\chi]$ represents
the Polyakov action including the dilaton fluctuation
\be
S_p[X,s,\chi] = - \frac{T}{2} \int d^2 \s \ \sqrt{- \g} \g^{\a\b} \pa_{\a} X^A 
\pa_{\b} X^B G_{AB} \ e^{\chi/2} + \cdots .
\ee
For the magnon case, $I_{\chi}[X, s ;y]$ becomes
\bea
I_{\chi}[X,s;y] = i \frac{3}{\pi^2} \int d^2 \s \ {\cal L}_{st} \times  \ls \frac{z}{z^2 + (x-y)^2 } \rs^4 ,
\eea
where ${\cal L}_{st}$ corresponds to the Polyakov action in the absence of the dilaton field, which is given in \eq{act:ph0}.
Inserting solutions obtained in
the previous section, the above integration is reduced to
\bea \la{int:I}
I_{\chi}[X,s;y] &=& i \frac{3 T}{2 \pi^2} \int_{-s/2}^{s/2} d\ta \int_{-L}^{L} d \s
\lb \k^2 + \frac{1}{b^2 -a^2} \ls 2 b^2 \n^2 \sin^2 \th - \frac{\n c}{a} (a^2 + b^2) \rs \rb \nn
&& \qquad \qquad \qquad \qquad \times  \ls \frac{z}{z^2 + (x-y)^2 } \rs^4 .
\eea
Notice that $\k = i \r$ at the saddle point and that the propagator of the massless field  depends
on $\ta$ only.  In terms of $z=\cos \th$, $I_{\chi}[X,s;y] $ can be rewritten as the combination
of the elliptic integrals
\be
I_{\chi}[X,s;y] = - i \frac{6 T}{ \pi^2}  \frac{\r \ls z_{max}^2 E(x) - z_{min}^2 K(x) \rs }{z_{max} \sqrt{1-z_{min}^2}}   \int_{-s/2}^{s/2} d\ta \ls \frac{z}{z^2 + (x-y)^2 } \rs^4
\ee
For $\k s \gg 1 $, performing the $\ta$-integration using the solution in \eq{sol:ads} gives
at the leading order
\be
 \int_{-s/2}^{s/2} d\ta \ls \frac{z}{z^2 + (x-y)^2 } \rs^4 = \frac{1}{12 i \r} \frac{x_f^4}{y^4 \
 (x_f-y)^4} .
\ee
Using these results, we can expand $I_{\chi}[X,s;y]$ up to $z_{min}^2$ order as
\be
I_{\chi}[X,s;y] = - \frac{T}{2 \pi^2}   \lb \sin \frac{p}{2} - \frac{J}{4 T} z_{min}^2 - \frac{1}{4} \sin \frac{p}{2} \ z_{min}^2  \rb  \frac{x_f^{ 4} }{y^4 (x_f-y)^4}. 
\ee
After substituting the above together with \eq{rel:zmin} into \eq{for:three}, we can
finally find the three-point correlation function between two magnons and one marginal operator
\be
\bra {\cal O}_m (0) {\cal O}_m (x_f)  {\cal D}_{\chi} (y) \ket = \frac{1 }{2 \pi^2}  \lb
 - T \sin \frac{p}{2} + \ls  4 J \sin^2 \frac{p}{2}  + 4 T \sin^3 \frac{p}{2}  \rs  \ e^{- \frac{J}{T \sin \frac{p}{2}}  } \rb \frac{1}{x_f^{2 E - 4} y^4 (x_f-y)^4} .
\ee
So the structure constant $a_{{\cal D}AA}$ in the string theory side reads off
\be	\la{res:coupling}
2 \pi^2 a_{{\cal D}mm} = 
 - T \sin \frac{p}{2} + \ls  4 J \sin^2 \frac{p}{2}  + 4 T \sin^3 \frac{p}{2}  \rs  \ e^{- \frac{J}{T \sin \frac{p}{2}}  } 
\ee
If taking $J \to \infty$, the above is reduced to the structure constant between two infinite
size magnons and one marginal operator. So the second term in \eq{res:coupling} corresponds
to the leading finite size correction in the large $J/T$ limit.

In the dual conformal field theory, the three-point correlation function between two magnons
and one marginal operator is given by
\be
\bra {\cal O}_m (0) {\cal O}_m (x_f)  {\cal D}_{\chi} (y) \ket = \frac{a_{{\cal D}mm} }{x_f^{2 E - 4} y^4 (x_f-y)^4} ,
\ee
where the denominator is fixed by the global conformal transformation.  The unknown
structure constant, which is not determined by the conformal symmetry, can be fixed by the 
formular obtained from the RG analysis 
\be
a_{{\cal D}mm}  = - g^2 \frac{\pa}{\pa g^2} \D = - \frac{T}{2} \frac{\pa}{\pa T} \D ,
\ee
where $\D$ is the conformal dimension of the magnon and $T = 2g$. 
Since at the large 't Hooft coupling regime the magnon's conformal dimension  
is the same as the energy of the giant magnon in \eq{rel:disp}, the structure constant between two
magnons having finite size and a marginal operator are given by
\bea 	\la{res:str}
a_{{\cal D}mm}  &=& - \frac{T}{2} \frac{\pa}{\pa T} \ls  J + 2 T \sin \frac{p}{2} - 8 T \sin^3 \frac{p}{2} \ e^{- \frac{J}{T \sin \frac{p}{2}}} \rs \nn
&=&  - T \sin \frac{p}{2} + \ls  4 J \sin^2 \frac{p}{2}  + 4 T \sin^3 \frac{p}{2}  \rs  \ e^{- \frac{J}{T \sin \frac{p}{2}}  } ,
\eea
which is the same as the result obtained from the string calculation.
Notice that though the angular momentum $J$ in the string theory is proportional to the string tension $T$, $J$ in ${\cal N}=4$ SYM corresponds to the number of scalar fields, so $J$
is independent of the coupling $g^2$, which means $\frac{\pa}{\pa T} J = 0$ in \eq{res:str}.

\section{Discussion}

We calculated the finite size correction on the two- and three-point correlation functions of 
the giant magnon. By the saddle point approximation, we rederived the finite size effect for the
the dispersion relation of the giant magnon. We also calculated the finite size effect
on the three point correlation function between two giant magnons and one marginal operator,
whose result is exactly the same as one obtained by the RG analysis. The calculation of the finite
size correction on the giant magnon can be easily extended to the dyonic magnon case moving
on $AdS_5 \times S^3$.

It is interesting to investigate the three point correlation function of heavy operators
like the giant magnon with the relevant or irrelevant light operator instead of the marginal
one. We hope to report these issues elsewhere.

\vspace{1cm}

{\bf Acknowledgement}

This work was supported by the National Research Foundation of Korea(NRF) grant funded by
the Korea government(MEST) through the Center for Quantum Spacetime(CQUeST) of Sogang
University with grant number 2005-0049409. C. Park was also
supported by Basic Science Research Program through the
National Research Foundation of Korea(NRF) funded by the Ministry of
Education, Science and Technology(2010-0022369).


\vspace{1cm}






\begin{thebibliography}{99}

\bibitem{mal1}
J. M. Maldacena, %"The Large N Limit of Superconformal Field Theories and Supergravity",
Adv. Theor. Math. Phys. {\bf 2}, 231, (1998); Int. J. Theor. Phys. {\bf 38}, 1113
(1999).

%\cite{Beisert:2005tm}
\bibitem{Beisert:2005tm}
  N.~Beisert,
  %``The su(2|2) dynamic S-matrix,''
  Adv.\ Theor.\ Math.\ Phys.\  {\bf 12}, 945 (2008)
  [arXiv:hep-th/0511082].
  %%CITATION = 00203,12,945;%%

%\cite{Minahan:2002ve}
\bibitem{Minahan:2002ve}
  J.~A.~Minahan and K.~Zarembo,
  %``The Bethe-ansatz for N = 4 super Yang-Mills,''
  JHEP {\bf 0303}, 013 (2003)
  [arXiv:hep-th/0212208].
  %%CITATION = JHEPA,0303,013;%%

%\cite{Beisert:2003xu}
\bibitem{Beisert:2003xu}
  N.~Beisert, J.~A.~Minahan, M.~Staudacher and K.~Zarembo,
  %``Stringing spins and spinning strings,''
  JHEP {\bf 0309}, 010 (2003)
  [arXiv:hep-th/0306139].
  %%CITATION = JHEPA,0309,010;%%

%\cite{Beisert:2003yb}
\bibitem{Beisert:2003yb}
  N.~Beisert and M.~Staudacher,
  %``The N=4 SYM Integrable Super Spin Chain,''
  Nucl.\ Phys.\  B {\bf 670}, 439 (2003)
  [arXiv:hep-th/0307042].
  %%CITATION = NUPHA,B670,439;%%

%\cite{Ishizeki:2007we}
\bibitem{Ishizeki:2007we}
  R.~Ishizeki and M.~Kruczenski,
  %``Single spike solutions for strings on S2 and S3,''
  Phys.\ Rev.\  D {\bf 76}, 126006 (2007)
  [arXiv:0705.2429 [hep-th]].
  %%CITATION = PHRVA,D76,126006;%%

%\cite{Chen:2006gp}
\bibitem{Chen:2006gp}
  H.~Y.~Chen, N.~Dorey and K.~Okamura,
  %``The asymptotic spectrum of the N = 4 super Yang-Mills spin chain,''
  JHEP {\bf 0703}, 005 (2007)
  [arXiv:hep-th/0610295].
  %%CITATION = JHEPA,0703,005;%%

%\cite{Hofman:2006xt}
\bibitem{Hofman:2006xt}
  D.~M.~Hofman and J.~M.~Maldacena,
  %``Giant magnons,''
  J.\ Phys.\ A  {\bf 39}, 13095 (2006)
  [arXiv:hep-th/0604135].
  %%CITATION = JPAGB,A39,13095;%%

%\cite{Frolov:2006cc}
\bibitem{Frolov:2006cc}
  S.~Frolov, J.~Plefka and M.~Zamaklar,
  %``The AdS(5) x S**5 superstring in light-cone gauge and its Bethe
  %equations,''
  J.\ Phys.\ A  {\bf 39}, 13037 (2006)
  [arXiv:hep-th/0603008].
  %%CITATION = JPAGB,A39,13037;%%

%\cite{Dorey:2008zz}
\bibitem{Dorey:2008zz}
  N.~Dorey,
  %``Integrability And The Ads/Cft Correspondence,''
  Class.\ Quant.\ Grav.\  {\bf 25}, 214003 (2008).
  %%CITATION = CQGRD,25,214003;%%

  %\cite{Arutyunov:2006gs}
\bibitem{Arutyunov:2006gs}
  G.~Arutyunov, S.~Frolov and M.~Zamaklar,
  %``Finite-size effects from giant magnons,''
  Nucl.\ Phys.\  B {\bf 778}, 1 (2007)
  [arXiv:hep-th/0606126].
  %%CITATION = NUPHA,B778,1;%%

%\cite{Lee:2008sk}
\bibitem{Lee:2008sk}
  B.~H.~Lee, R.~R.~Nayak, K.~L.~Panigrahi and C.~Park,
  %``On the giant magnon and spike solutions for strings on AdS$_3\times$
  %S$^3$,''
  JHEP {\bf 0806}, 065 (2008)
  [arXiv:0804.2923 [hep-th]].
  %%CITATION = JHEPA,0806,065;%%

%\cite{Kluson:2008gf}
\bibitem{Kluson:2008gf}
  J.~Kluson, B.~H.~Lee, K.~L.~Panigrahi and C.~Park,
  %``Magnon like solutions for strings in I-brane background,''
  JHEP {\bf 0808}, 032 (2008)
  [arXiv:0806.3879 [hep-th]].
  %%CITATION = JHEPA,0808,032;%%

%\cite{Gwak:2009nq}
\bibitem{Gwak:2009nq}
  B.~Gwak, B.~H.~Lee, K.~L.~Panigrahi and C.~Park,
  %``Semiclassical strings in AdS(3) X S^2,''
  JHEP {\bf 0904}, 071 (2009)
  [arXiv:0901.2795 [hep-th]].
  %%CITATION = JHEPA,0904,071;%%

%\cite{Okamura:2008jm}
\bibitem{Okamura:2008jm}
  K.~Okamura,
  %``Aspects of Integrability in AdS/CFT Duality,''
  arXiv:0803.3999 [hep-th].
  %%CITATION = ARXIV:0803.3999;%%

%\cite{Arutyunov:2004vx}
\bibitem{Arutyunov:2004vx}
  G.~Arutyunov, S.~Frolov and M.~Staudacher,
  %``Bethe ansatz for quantum strings,''
  JHEP {\bf 0410}, 016 (2004)
  [arXiv:hep-th/0406256].
  %%CITATION = JHEPA,0410,016;%%

  %\cite{Arutyunov:2007tc}
\bibitem{Arutyunov:2007tc}
  G.~Arutyunov and S.~Frolov,
  %``On String S-matrix, Bound States and TBA,''
  JHEP {\bf 0712}, 024 (2007)
  [arXiv:0710.1568 [hep-th]].
  %%CITATION = JHEPA,0712,024;%%
  
  %\cite{Arutyunov:2009zu}
\bibitem{Arutyunov:2009zu}
  G.~Arutyunov and S.~Frolov,
  %``String hypothesis for the AdS(5) x S**5 mirror,''
  JHEP {\bf 0903}, 152 (2009)
  [arXiv:0901.1417 [hep-th]].
  %%CITATION = JHEPA,0903,152;%%

%\cite{Gromov:2009tv}
\bibitem{Gromov:2009tv}
  N.~Gromov, V.~Kazakov and P.~Vieira,
  %``Exact Spectrum of Anomalous Dimensions of Planar N=4 Supersymmetric
  %Yang-Mills Theory,''
  Phys.\ Rev.\ Lett.\  {\bf 103}, 131601 (2009)
  [arXiv:0901.3753 [hep-th]].
  %%CITATION = PRLTA,103,131601;%%

%\cite{Arutyunov:2009ur}
\bibitem{Arutyunov:2009ur}
  G.~Arutyunov and S.~Frolov,
  %``Thermodynamic Bethe Ansatz for the AdS(5) x S(5) Mirror Model,''
  JHEP {\bf 0905}, 068 (2009)
  [arXiv:0903.0141 [hep-th]].
  %%CITATION = JHEPA,0905,068;%%

%\cite{Arutyunov:2009kf}
\bibitem{Arutyunov:2009kf}
  G.~Arutyunov and S.~Frolov,
  %``The Dressing Factor and Crossing Equations,''
  J.\ Phys.\ A  {\bf 42}, 425401 (2009)
  [arXiv:0904.4575 [hep-th]].
  %%CITATION = JPAGB,A42,425401;%%
 
 %\cite{Bombardelli:2009ns}
\bibitem{Bombardelli:2009ns}
  D.~Bombardelli, D.~Fioravanti and R.~Tateo,
  %``Thermodynamic Bethe Ansatz for planar AdS/CFT: A Proposal,''
  J.\ Phys.\ A  {\bf 42}, 375401 (2009)
  [arXiv:0902.3930 [hep-th]].
  %%CITATION = JPAGB,A42,375401;%%
   
%\cite{Gromov:2009bc}
\bibitem{Gromov:2009bc}
  N.~Gromov, V.~Kazakov, A.~Kozak and P.~Vieira,
  %``Exact Spectrum of Anomalous Dimensions of Planar N = 4 Supersymmetric
  %Yang-Mills Theory: TBA and excited states,''
  Lett.\ Math.\ Phys.\  {\bf 91}, 265 (2010)
  [arXiv:0902.4458 [hep-th]].
  %%CITATION = LMPHD,91,265;%%

%\cite{Cavaglia:2010nm}
\bibitem{Cavaglia:2010nm}
  A.~Cavaglia, D.~Fioravanti and R.~Tateo,
  %``Extended Y-system for the $AdS_5/CFT_4$ correspondence,''
  Nucl.\ Phys.\  B {\bf 843}, 302 (2011)
  [arXiv:1005.3016 [hep-th]].
  %%CITATION = NUPHA,B843,302;%%
  
  %\cite{Cavaglia:2011kd}
\bibitem{Cavaglia:2011kd}
  A.~Cavaglia, D.~Fioravanti, M.~Mattelliano and R.~Tateo,
  %``On the $AdS_5/CFT_4$ TBA and its analytic properties,''
  arXiv:1103.0499 [hep-th].
  %%CITATION = ARXIV:1103.0499;%%
  
  

 
  
%\cite{Aharony:2008ug}
\bibitem{Aharony:2008ug}
  O.~Aharony, O.~Bergman, D.~L.~Jafferis and J.~Maldacena,
  %``N=6 superconformal Chern-Simons-matter theories, M2-branes and their
  %gravity duals,''
  JHEP {\bf 0810}, 091 (2008)
  [arXiv:0806.1218 [hep-th]].
  %%CITATION = JHEPA,0810,091;%%

%\cite{Grignani:2008is}
\bibitem{Grignani:2008is}
  G.~Grignani, T.~Harmark and M.~Orselli,
  %``The SU(2) x SU(2) sector in the string dual of N=6 superconformal
  %Chern-Simons theory,''
  Nucl.\ Phys.\  B {\bf 810}, 115 (2009)
  [arXiv:0806.4959 [hep-th]].
  %%CITATION = NUPHA,B810,115;%%

%\cite{Grignani:2008te}
\bibitem{Grignani:2008te}
  G.~Grignani, T.~Harmark, M.~Orselli and G.~W.~Semenoff,
  %``Finite size Giant Magnons in the string dual of N=6 superconformal
  %Chern-Simons theory,''
  JHEP {\bf 0812}, 008 (2008)
  [arXiv:0807.0205 [hep-th]].
  %%CITATION = JHEPA,0812,008;%%
  
  %\cite{Astolfi:2008ji}
\bibitem{Astolfi:2008ji}
  D.~Astolfi, V.~G.~M.~Puletti, G.~Grignani, T.~Harmark and M.~Orselli,
  %``Finite-size corrections in the SU(2) x SU(2) sector of type IIA string
  %theory on AdS_4 x CP^3,''
  Nucl.\ Phys.\  B {\bf 810}, 150 (2009)
  [arXiv:0807.1527 [hep-th]].
  %%CITATION = NUPHA,B810,150;%%
  
  %\cite{Astolfi:2009qh}
\bibitem{Astolfi:2009qh}
  D.~Astolfi, V.~G.~M.~Puletti, G.~Grignani, T.~Harmark and M.~Orselli,
  %``Full Lagrangian and Hamiltonian for quantum strings on AdS_4 x CP^3 in a
  %near plane wave limit,''
  JHEP {\bf 1004}, 079 (2010)
  [arXiv:0912.2257 [hep-th]].
  %%CITATION = JHEPA,1004,079;%%
 

  %\cite{Shenderovich:2008bs}
\bibitem{Shenderovich:2008bs}
  I.~Shenderovich,
  %``Giant magnons in AdS_4/CFT_3: dispersion, quantization and finite--size
  %corrections,''
  arXiv:0807.2861 [hep-th].
  %%CITATION = ARXIV:0807.2861;%%

%\cite{McLoughlin:2008ms}
\bibitem{McLoughlin:2008ms}
  T.~McLoughlin and R.~Roiban,
  %``Spinning strings at one-loop in AdS_4 x P^3,''
  JHEP {\bf 0812}, 101 (2008)
  [arXiv:0807.3965 [hep-th]].
  %%CITATION = JHEPA,0812,101;%%

%\cite{Alday:2008ut}
\bibitem{Alday:2008ut}
  L.~F.~Alday, G.~Arutyunov and D.~Bykov,
  %``Semiclassical Quantization of Spinning Strings in AdS_4 x CP^3,''
  JHEP {\bf 0811}, 089 (2008)
  [arXiv:0807.4400 [hep-th]].
  %%CITATION = JHEPA,0811,089;%%

%\cite{Lee:2008ui}
\bibitem{Lee:2008ui}
  B.~H.~Lee, K.~L.~Panigrahi and C.~Park,
  %``Spiky Strings on AdS$_4 \times {\bf CP}^3$,''
  JHEP {\bf 0811}, 066 (2008)
  [arXiv:0807.2559 [hep-th]].
  %%CITATION = JHEPA,0811,066;%%

%\cite{Baek:2008ws}
\bibitem{Baek:2008ws}
  J.~H.~Baek, S.~Hyun, W.~Jang and S.~H.~Yi,
  %``Membrane Dynamics in Three dimensional N=6 Supersymmetric Chern-Simons
  %Theory,''
  arXiv:0812.1772 [hep-th].
  %%CITATION = ARXIV:0812.1772;%%

%\cite{Lee:2008yq}
\bibitem{Lee:2008yq}
  B.~H.~Lee and C.~Park,
  %``Unbounded Multi Magnon and Spike,''
J. Korean Phys.Soc. {\bf 57} 30  (2010)
 [arXiv:0812.2727 [hep-th]].
  %%CITATION = ARXIV:0812.2727;%%

%\cite{Kalousios:2009mp}
\bibitem{Kalousios:2009mp}
  C.~Kalousios, M.~Spradlin and A.~Volovich,
  %``Dyonic Giant Magnons on CP^3,''
  JHEP {\bf 0907}, 006 (2009)
  [arXiv:0902.3179 [hep-th]].
  %%CITATION = JHEPA,0907,006;%%

%\cite{Ahn:2008aa}
\bibitem{Ahn:2008aa}
  C.~Ahn and R.~I.~Nepomechie,
  %``N=6 super Chern-Simons theory S-matrix and all-loop Bethe ansatz
  %equations,''
  JHEP {\bf 0809}, 010 (2008)
  [arXiv:0807.1924 [hep-th]].
  %%CITATION = JHEPA,0809,010;%%

%\cite{Jevicki:2009uz}
\bibitem{Jevicki:2009uz}
  A.~Jevicki and K.~Jin,
  %``Moduli Dynamics of AdS_3 Strings,''
  JHEP {\bf 0906}, 064 (2009)
  [arXiv:0903.3389 [hep-th]].
  %%CITATION = JHEPA,0906,064;%%

%\cite{Minahan:2009te}
\bibitem{Minahan:2009te}
  J.~A.~Minahan, W.~Schulgin and K.~Zarembo,
  %``Two loop integrability for Chern-Simons theories with N=6 supersymmetry,''
  JHEP {\bf 0903} (2009) 057
  [arXiv:0901.1142 [hep-th]].
  %%CITATION = JHEPA,0903,057;%%

%\cite{Abbott:2009um}
\bibitem{Abbott:2009um}
  M.~C.~Abbott, I.~Aniceto and O.~Ohlsson Sax,
  %``Dyonic Giant Magnons in CP^3: Strings and Curves at Finite J,''
  Phys.\ Rev.\  D {\bf 80}, 026005 (2009)
  [arXiv:0903.3365 [hep-th]].
  %%CITATION = PHRVA,D80,026005;%%

%\cite{Bombardelli:2009xz}
\bibitem{Bombardelli:2009xz}
  D.~Bombardelli, D.~Fioravanti and R.~Tateo,
  %``TBA and Y-system for planar AdS(4)/CFT(3),''
  Nucl.\ Phys.\  B {\bf 834}, 543 (2010)
  [arXiv:0912.4715 [hep-th]].
  %%CITATION = NUPHA,B834,543;%%

%\cite{Gromov:2009at}
\bibitem{Gromov:2009at}
  N.~Gromov and F.~Levkovich-Maslyuk,
  %``Y-system, TBA and Quasi-Classical strings in AdS(4) x CP3,''
  JHEP {\bf 1006}, 088 (2010)
  [arXiv:0912.4911 [hep-th]].
  %%CITATION = JHEPA,1006,088;%%
  



%\cite{Janik:2010gc}
\bibitem{Janik:2010gc}
  R.~A.~Janik, P.~Surowka and A.~Wereszczynski,
  %``On correlation functions of operators dual to classical spinning string
  %states,''
  JHEP {\bf 1005}, 030 (2010)
  [arXiv:1002.4613 [hep-th]].
  %%CITATION = JHEPA,1005,030;%%

%\cite{Buchbinder:2010vw}
\bibitem{Buchbinder:2010vw}
  E.~I.~Buchbinder and A.~A.~Tseytlin,
  %``On semiclassical approximation for correlators of closed string vertex
  %operators in AdS/CFT,''
  JHEP {\bf 1008}, 057 (2010)
  [arXiv:1005.4516 [hep-th]].
  %%CITATION = JHEPA,1008,057;%%

%\cite{Grossardt:2010xq}
\bibitem{Grossardt:2010xq}
  A.~Grossardt and J.~Plefka,
  %``One-Loop Spectroscopy of Scalar Three-Point Functions in planar N=4 super
  %Yang-Mills Theory,''
  arXiv:1007.2356 [hep-th].
  %%CITATION = ARXIV:1007.2356;%%

%\cite{Zarembo:2010rr}
\bibitem{Zarembo:2010rr}
  K.~Zarembo,
  %``Holographic three-point functions of semiclassical states,''
  JHEP {\bf 1009}, 030 (2010)
  [arXiv:1008.1059 [hep-th]].
  %%CITATION = JHEPA,1009,030;%%

%\cite{Costa:2010rz}
\bibitem{Costa:2010rz}
  M.~S.~Costa, R.~Monteiro, J.~E.~Santos and D.~Zoakos,
  %``On three-point correlation functions in the gauge/gravity duality,''
  arXiv:1008.1070 [hep-th].
  %%CITATION = ARXIV:1008.1070;%%

%\cite{Roiban:2010fe}
\bibitem{Roiban:2010fe}
  R.~Roiban and A.~A.~Tseytlin,
  %``On semiclassical computation of 3-point functions of closed string vertex
  %operators in AdS_5 x S^5,''
  arXiv:1008.4921 [hep-th].
  %%CITATION = ARXIV:1008.4921;%%

%\cite{Hernandez:2010tg}
\bibitem{Hernandez:2010tg}
  R.~Hernandez,
  %``Three-point correlation functions from semiclassical circular strings,''
  arXiv:1011.0408 [hep-th].
  %%CITATION = ARXIV:1011.0408;%%

%\cite{Ryang:2010bn}
\bibitem{Ryang:2010bn}
  S.~Ryang,
  %``Correlators of Vertex Operators for Circular Strings with Winding Numbers
  %in AdS5xS5,''
  arXiv:1011.3573 [hep-th].
  %%CITATION = ARXIV:1011.3573;%%

%\cite{Bak:2011yy}
\bibitem{Bak:2011yy}
  D.~Bak, B.~Chen and J.~B.~Wu,
  %``Holographic Correlation Functions for Open Strings and Branes,''
  arXiv:1103.2024 [hep-th].
  %%CITATION = ARXIV:1103.2024;%%
  
%\cite{Bissi:2011dc}
\bibitem{Bissi:2011dc}
  A.~Bissi, C.~Kristjansen, D.~Young and K.~Zoubos,
  %``Holographic three-point functions of giant gravitons,''
  arXiv:1103.4079 [hep-th].
  %%CITATION = ARXIV:1103.4079;%%
  
%\cite{Arnaudov:2010kk}
\bibitem{Arnaudov:2010kk}
  D.~Arnaudov and R.~C.~Rashkov,
  %``On semiclassical calculation of three-point functions in AdS_4 x CP^3,''
  arXiv:1011.4669 [hep-th].
  %%CITATION = ARXIV:1011.4669;%%

%\cite{Georgiou:2010an}
\bibitem{Georgiou:2010an}
  G.~Georgiou,
  %``Two and three-point correlators of operators dual to folded string
  %solutions at strong coupling,''
  arXiv:1011.5181 [hep-th].
  %%CITATION = ARXIV:1011.5181;%%

%\cite{Hernandez:2011up}
\bibitem{Hernandez:2011up}
  R.~Hernandez,
  %``Three-point correlators for giant magnons,''
  arXiv:1104.1160 [hep-th].
  %%CITATION = ARXIV:1104.1160;%%
  
%\cite{Janik:2006dc}
\bibitem{Janik:2006dc}
  R.~A.~Janik,
  %``The AdS(5) x S**5 superstring worldsheet S-matrix and crossing symmetry,''
  Phys.\ Rev.\  D {\bf 73}, 086006 (2006)
  [arXiv:hep-th/0603038].
  %%CITATION = PHRVA,D73,086006;%%
  
%\cite{Park:2010vs}
\bibitem{Park:2010vs}
  C.~Park and B.~H.~Lee,
  %``Correlation functions of magnon and spike,''
  arXiv:1012.3293 [hep-th].
  %%CITATION = ARXIV:1012.3293;%%
  
  %\cite{Bai:2011su}
\bibitem{Bai:2011su}
  X.~Bai, B.~H.~Lee and C.~Park,
  %``Correlation function of dyonic strings,''
  arXiv:1104.1896 [hep-th].
  %%CITATION = ARXIV:1104.1896;%%
  
  %\cite{Ahn:2011zg}
\bibitem{Ahn:2011zg}
  C.~Ahn and P.~Bozhilov,
  %``Three-point Correlation functions of Giant magnons with finite size,''
  arXiv:1105.3084 [hep-th].
  %%CITATION = ARXIV:1105.3084;%%
  
  %\cite{Freedman:1998tz}
\bibitem{Freedman:1998tz}
  D.~Z.~Freedman, S.~D.~Mathur, A.~Matusis, L.~Rastelli,
  %``Correlation functions in the CFT(d) / AdS(d+1) correspondence,''
  Nucl.\ Phys.\  {\bf B546}, 96-118 (1999).
  [hep-th/9804058].

\end{thebibliography}
\end{document}